\documentclass[12pt]{article}
\usepackage{graphicx}
\usepackage{amsfonts}
\usepackage{amssymb,amsmath}
\usepackage{latexsym}
\usepackage{color}
\usepackage{cite}
\input{colordvi.tex}

\begin{document}

\begin{titlepage}

\begin{flushright}
IPMU 09-0042 \\
ICRR-Report-540
\end{flushright}

\begin{center}

{\Large \bf  
Constraining Light Gravitino Mass from \vspace{0.2cm} \\
Cosmic Microwave Background}

\vskip .45in

{\large
Kazuhide Ichikawa$^{1}$, 
Masahiro Kawasaki$^{2,3}$, 
Kazunori Nakayama$^2$, \\
\vspace{0.3cm}
Toyokazu Sekiguchi$^2$
and 
Tomo Takahashi$^4$
}

\vskip .45in

{\em
$^1$Department of Micro Engineering, Kyoto University, 
Kyoto 606-8501, Japan \vspace{0.2cm} \\
$^2$Institute for Cosmic Ray Research, University of Tokyo,
Kashiwa 277-8582, Japan  \vspace{0.2cm} \\
$^3$Institute for the Physics and Mathematics of the Universe,
University of Tokyo, Kashiwa, Chiba, 277-8568, Japan \vspace{0.2cm}\\
$^4$Department of Physics, Saga University, Saga 840-8502, Japan \\
}

\end{center}

\vskip .4in

\begin{abstract}
 We investigate the
  possibilities of constraining the light gravitino mass $m_{3/2}$
  from future cosmic microwave background (CMB) surveys.
  A model with light gravitino with the mass $m_{3/2} < \mathcal{O}(10) \ \mbox{eV}$
  is of great interest since it is free from the cosmological gravitino problem
 and, in addition, can be compatible with many baryogenesis/leptogenesis 
 scenarios such as the thermal leptogenesis.   We show that
  the lensing of CMB anisotropies can be a good probe for $m_{3/2}$
  and obtain an expected constraint on $m_{3/2}$ from 
  precise measurements of lensing potential in the future CMB
  surveys, such as the PolarBeaR and CMBpol experiments. 
  If the gravitino mass is $m_{3/2} = 1 \ \mbox{eV}$, we will obtain the constraint for 
  the gravitino mass as $m_{3/2}\le 3.2 \ \mbox{eV} \
\mbox{(95\% C.L.)}$ for the case with Planck+PolarBeaR combined and
$m_{3/2}=1.04^{+0.22}_{-0.26} \ \mbox{eV} \ \mbox{(68\% C.L.)}$ for
CMBpol.  The issue of Bayesian model selection is also discussed.
	
\end{abstract}

\end{titlepage}

\setcounter{page}{1}

\section{Introduction} \label{sec:introduction}

One of the most important prediction of local supersymmetry (SUSY), or
supergravity, is the existence of gravitino, the spin-$3/2$
superpartner of the graviton.  Although the range of the gravitino
mass $m_{3/2}$ can vary from a fraction of eV up to of an order of
TeV, depending on the scales of SUSY breaking, a light gravitino with
$m_{3/2} < \mathcal{O}(10)$ eV is of great interest since it is free
from the cosmological gravitino problem.  Furthermore, for some
baryogenesis scenario to work such as thermal leptgenesis, high cosmic
temperature is required, which favors the range of light mass for the
gravitino.  Thus, in this respect, the light gravitino would be
attractive.

The determination of the gravitino mass is one of the most important
issues to understand how supersymmetry is broken.  Some authors have
discussed this issue, in particular, focusing on probing the mass of
the light gravitino with LHC experiment \cite{Hamaguchi:2007ge}.  
Since gravitino with very light mass of $m_{3/2} < 100 \ \mbox{eV}$ can play 
a role of warm dark matter (WDM) 
in the universe, cosmology would be a powerful tool as well.
Some authors have obtained a constraint on the light gravitino mass
from Lyman-$\alpha$ forest data in combination with WMAP
\cite{Viel:2005qj,Boyarsky:2008xj} and its bound is $m_{3/2}<16~$ eV at 2$\sigma$
level \cite{Viel:2005qj}\footnote{
Recently the authors of  \cite{Boyarsky:2008xj} also 
conducted a similar analysis using Lyman-$\alpha$ data
combined with updated WMAP5~\cite{Komatsu:2008hk}.
Although they did not report the constraint on the light gravitino mass,
the SDSS Lyman-$\alpha$ dataset would be more 
effective in constraining $m_{3/2}$ than the dataset adopted in \cite{Viel:2005qj}.
Thus an analysis optimized for the light gravitino model 
could give a severer constraint on $m_{3/2}$.
}.
Although Lyman-$\alpha$ forest can be very useful as a cosmological probe, 
it may suffer from some systematic uncertainties \cite{Viel:2006yh,Seljak:2006bg}.
Notice that the above mentioned limit crucially
depends on the usage of Lyman-$\alpha$ forest data, thus in this
respect, another independent cosmological probe of the light gravitino
would be of great importance.  Since light gravitino with the mass of
interest here acts as WDM, it affects cosmic density
fluctuations through two major effects.  One is the change of
radiation-matter equality due to the fact that light gravitions behave as
relativistic component at earlier times.  The second one is the effect
of free-streaming which erases density contrast on the scale under
which it can free-stream.  Since these two effects can make great influence
on cosmic microwave background (CMB) anisotropy, CMB can be a
powerful probe of its mass.  But in fact, when the gravitino is very
light as $m_{3/2} < \mathcal{O}(10)$ eV, which is the range of the
mass of interest here, its energy density do not have a large fraction
of the total one in general to affect the CMB.  Thus the mass of light gravitino
can be mainly probed through the latter effect, free-streaming.  
The ultralight gravitino is almost
relativistic at the time of the recombination, its effects imprinted
on temperature anisotropy is not significant enough.  However, by
looking at the gravitational lensing of CMB photons, we can well probe
the change of the gravitational potential driven during the
intermediate redshift after the recombination when light gravitino
comes to act as a non-relativistic component.  Thus the lensed CMB is
a very useful cosmological tool for investigating the mass of light
gravitino\footnote{
A similar argument has been made for massive neutrinos
on the use of CMB lensing to constrain its mass Ref.~\cite{Kaplinghat:2003bh,Lesgourgues:2005yv,Perotto:2006rj}.
}, which we discuss in this paper.  Since
much more precise CMB experiments will be available in the near
future, it is an interesting subject to investigate possible limits
with such a future probe.  Therefore, we in this paper study the possibilities
of constraining its mass with future CMB observations without
using other cosmological data. In particular, we focus on the lensing
potential which can be reconstracted from CMB maps.  As future CMB surveys,
we consider Planck, PolarBeaR and CMBpol to discuss
a possible constraint on the light gravitino mass.

The structure of this paper is as follows.  We first briefly review a
model which predicts light gravitino and its phenomenology in the
early universe in Section~\ref{sec:model}. In Section~\ref{sec:CMB} we
discuss the effects of the light gravitino on CMB anisotropy, 
paying particular attention to the lensing of CMB. We then present
forecasts for constraints on the mass of light gravitino with future
CMB surveys such as Planck, PolarBeaR and CMBpol in
Section~\ref{sec:constraints}.  
In addition to the parameter estimation, we also discuss 
Bayesian model selection analysis for light gravitino model 
with future CMB surveys in Section~\ref{sec:selection}.
The final section is devoted to
summary of this paper.

\section{Light gravitino: A model and its phenomenology in the early universe
}\label{sec:model}

A light gravitino scenario is realized in the framework of
gauge-mediated SUSY breaking (GMSB) models \cite{Giudice:1998bp}.  In
GMSB models, the SUSY breaking effect in the hidden sector is
transmitted to the minimal supersymmetric standard model (MSSM) sector
through gauge-interactions, giving superparticles TeV scale masses.
As an example, let us consider a model where the SUSY breaking field
$S$ couples to $N$ pairs of messenger particles $\psi$ and $\bar
\psi$, which transform as fundamental and anti-fundamental
representations of SU(5), having a superpotential $W = \lambda S \psi
\bar \psi$ with coupling constant $\lambda$ (in the following we set
$\lambda = 1$ for simplicity).  The superfield $S$ has a vacuum
expectation value as $\langle S \rangle = M+F_S \theta^2$.  Here $F_S$
gives SUSY breaking scale, which is related to the garvitino mass
$m_{3/2}$ through the relation $F_S = \sqrt 3 m_{3/2} M_P$, with $M_P$ being 
the reduced Planck energy scale,  for
vanishing cosmological constant.  In this model, gaugino masses $M_a$
($a=1,2,3$ are gauge indices) and sfermion masses squared $m_{\tilde
  f_i}^2$ at the messenger scale are given by
\begin{gather}
	M_a =N\left (\frac{\alpha_a}{4\pi}\right) \Lambda_{\rm mess}, \\
	m_{\tilde f_i}^2 = 2N\sum_a \left ( \frac{\alpha_a}{4\pi} \right )^2C_a^{(i)}
	\Lambda_{\rm mess}^2,
\end{gather}
where $\alpha_a$ denotes the gauge coupling constants, $C_a^{(i)}$ are
Casimir operators for the sfermion $\tilde f_i$ and the messenger
scale is given by $\Lambda_{\rm mess} = F_S/M$.  In order to obtain
TeV scale masses, $\Lambda_{\rm mess} \sim 100$~TeV is required, but
still the SUSY breaking scale $F_S$, or gravitino mass $m_{3/2}$ can
take wide range of values as $1~{\rm eV}\lesssim m_{3/2} \lesssim
10$~GeV.  The upper bound comes from the requirement that the
gravity-mediation effect does not dominate.  On the other hand, there
also exists a lower bound on the gravitino mass in order not to
destabilize the messenger scalar and lead to the unwanted vacuum.
This requires $m_{3/2} \gtrsim \mathcal O(1)$ eV.

However, if cosmological effects of the gravitino are taken into
account, not all of its mass range is favored.  This is because
gravitinos are efficiently produced at the reheating era and it can
easily exceed the present dark matter abundance unless the reheating
temperature $T_R$ is very low \cite{Moroi:1993mb}.  This is problematic
since many known leptogenesis/baryogenesis scenarios require high
enough reheating temperature which may conflict with the upper bound
coming from the gravitino problem.  In particular, thermal
leptogenesis scenario \cite{Fukugita:1986hr}, which requires $T_R
\gtrsim 10^9~$GeV, seems to be inconsistent with the gravitino problem
except for the very light gravitino mass range $m_{3/2} \lesssim
100$~eV.  As we will see, gravitinos with such a small mass are
thermalized in the early Universe.  Thus their abundance does not
depend on the reheating temperature and also it is smaller than the
dark matter density for $m_{3/2}\lesssim 100$~eV.  This is the reason
why we pay particular attention to a light gravitino scenario.

Having described that the light gravitino scenario is appealing from
the view point of cosmological gravitino problem,  next we briefly
discuss thermal evolution of the light gravitino in the early
universe.
Gravitinos are relativistic well before the recombination.  In such a case, 
the energy density of the gravitino is parameterized by the
effective number of neutrino species, and it is given by
\begin{equation}
	N_{3/2}=\frac{\rho_{3/2}}{\rho_\nu}=\left(\frac{T_{3/2}}{T_\nu} \right)^4=
\left(\frac{g_{*\nu}}{g_{*3/2}}\right)^{4/3},\label{eq:Ng}
\end{equation}
where $\rho_\nu$ ($\rho_{3/2}$) and $g_{*\nu}$ ($g_{*3/2}$) are,
respectively, the energy density and the effective degrees of freedom
of neutrinos (gravitinos) evaluated at the epoch when
neutrinos (gravitinos) have decoupled from thermal plasma while they are
still relativistic.  In the standard cosmology, $g_{*\nu}=10.75$.
Temperatures of neutrino and gravitino are represented by $T_\nu$ and
$T_{3/2}$.  From Eq.~(\ref{eq:Ng}) we can calculate the temperature of
gravitino at present:
\begin{equation}
	T_{3/2}= ( N_{3/2})^{1/4}T_\nu=1.95 (N_{3/2})^{1/4}~{\rm [K]},
\end{equation}
where we have adopted the temperature of neutrino in the standard
cosmology at the second equality.  Eventually the gravitino loses its
energy and becomes non-relativistic due to the Hubble expansion.  Its
present energy density is given by
\begin{equation}
	\label{eq:omega32}
	\omega_{3/2} \equiv \Omega_{3/2}h^2=
	0.13\left ( \frac{m_{3/2}}{100~{\rm eV}} \right )\left ( \frac{90}{g_{*3/2}} \right ).
\end{equation}
For later convenience, we also define 
the fraction
of gravitino in the total dark matter density $\omega_{\rm dm}$ as
\begin{equation}
f_{3/2} \equiv \frac{\omega_{3/2}}{\omega_{\rm dm}}.
\end{equation}
In the following, we assume that 
dark matter consists two components: light gravitino, which acts as 
warm dark matter, and CDM. As CDM component,  
the Peccei-Quinn axion, a messenger baryon proposed in 
\cite{Hamaguchi:2007rb} and so on 
can be well-fitted into the framework of light gravitino.

Thus in order to evaluate the relic abundance of light gravitino, we
must know the value of the effective degrees of freedom of
relativistic particles at the freeeze-out epoch, $g_{*3/2}$.  Since
the production and/or destruction of the light gravitino due to
scattering processes are known to be inefficient for the low
temperature regime, in which we are interested, the gravitino
maintains equilibrium with thermal bath through the decay and
inverse-decay processes \cite{Moroi:1993mb,Pierpaoli:1997im},
schematically represented by $a \leftrightarrow b +\tilde G$, where
$b$ is the standard model (SM) particle and $a$ is its superpartner.  As the temperature
decreases, particles in thermal bath ($b$ and $\tilde G$) lose an
ability to create a heavy particle ($a$).  Then gravitinos decouple
from thermal plasma after the time when $a$ decays into $b$ and
$\tilde G$ without inverse creation processes.

\begin{figure}[htb]
\begin{center}
  \scalebox{1.5}{\includegraphics{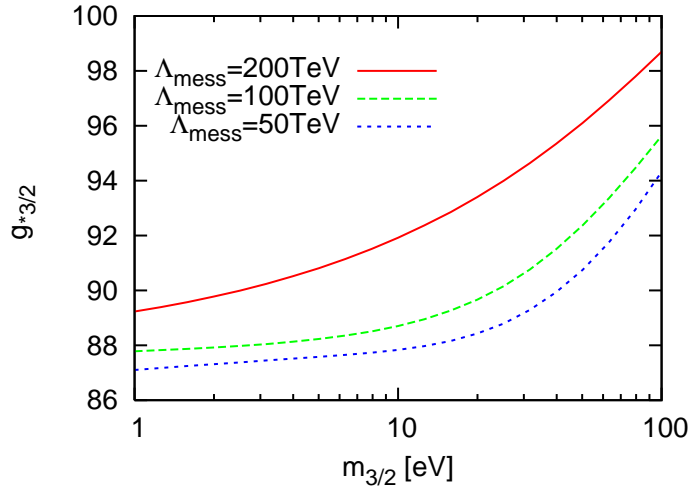}}
  \caption{$g_{*3/2}$ as a function of $m_{3/2}$ for several values of 
  $\Lambda_{\rm mess}$.
  }
\label{fig:g}
\end{center}
\end{figure}

In order to see these processes in detail, we must solve the Boltzmann
equation which governs time evolution of the system.  The Boltzmann
equation for the gravitino number density $n_{3/2}$ is given by
\begin{equation}
	\dot n_{3/2} +3Hn_{3/2} = \sum_{a,b}\Gamma (a \to b \tilde G)
	\left \langle \frac{m_a}{E_a} \right \rangle n_a
	\left (1- \frac{n_{3/2}}{n_{3/2}^{\rm (eq)}} \right ),
\end{equation}
where $H$ is the Hubble expansion rate and 
$\langle m_a/E_a \rangle$ represents thermally averaged Lorentz
factor with $m_a$ and $E_a$ being the mass and energy of the particle
$a$, respectively.  $\Gamma (a \to b \tilde G)$ is the decay width of
$a$ into $b$ and $\tilde G$ and superscript ${\rm (eq)}$ denotes its
equilibrium value.  As an example, the decay rate of the stau $(\tilde
\tau)$ into tau ($\tau$) and gravitino is given by
\begin{equation}
	\Gamma (\tilde \tau \to \tau \tilde G)=\frac{1}{48\pi}
	\frac{m_{\tilde \tau}^5}{m_{3/2}^2 M_P^2},
\end{equation}
and similar expressions hold for other particles.  By solving this
equation, one obtains the final gravitino abundance which can be
represented in terms of the gravitino-to-entropy ratio, $Y_{3/2}=
(n_{3/2}/s) (t\to \infty)$ and this translates into $g_{*3/2}$ through
the relation $Y_{3/2} = 0.417/g_{*3/2}$.  In Fig.~\ref{fig:g}, the
value of $g_{\ast 3/2}$ is shown as a function of $m_{3/2}$ for
several values of $\Lambda_{\rm mess}$.  We have adopted $\Lambda_{\rm
  mess}=50,100,200~{\rm TeV}$ and $N$=1 and ignored running of the masses from
the messenger scale down to the weak scale for simplicity.

We can understand these results intuitively.  As the gravitino mass
increases, the decay width becomes smaller, and hence the equilibrium
lasts for rather shorter duration.  This leads to higher freeze-out
temperature of the gravitino, which corresponds to large $g_{*3/2}$.
On the other hand, larger $\Lambda_{\rm mess}$ leads to heavier
sparticle masses, which obviously makes the time of freeze-out of the
gravitino earlier, and hence higher $g_{*3/2}$.  However, as seen in
the figure, the dependence on these parameters are not so strong and
we can safely set $g_{*3/2}\simeq 90$ even if we take into account
model uncertainties.

\section{Effects of light gravitino on CMB}\label{sec:CMB}

In this section we discuss how light gravitino affects the structure
formation and the CMB anisotropies. Since light gravitino basically
acts as WDM, it is characterized by two quantities,
its mass and number density.  The number density is determined by 
the effective number of degrees of freedom at the
time of the decoupling, i.e. $g_{*3/2}$.  Since, as we have seen in
the previous section, $g_{*3/2}$ has only mild dependence on
$m_{3/2}$, we can take $g_{*3/2}=90$ as the representative value,
which corresponds to
\begin{equation}
N_{3/2} = 0.059, \label{eq:fidN}
\end{equation}
from Eq.~(\ref{eq:Ng}).  We assume
Eq.~(\ref{eq:fidN}) as a fiducial value throughout this section, 
except for the last paragraph. We also assume that the
universe is flat, dark energy is a cosmological constant and the
primordial fluctuations are adiabatic and its power spectrum obeys a
power-law without tensor perturbations.  The fiducial values for
cosmological parameters are adopted from the the recent result of
WMAP5~\cite{Komatsu:2008hk}, except that we consider mixed dark matter
scenarios $\omega_{\rm dm} = \omega_c+\omega_{3/2}=0.1099$, instead of
$\omega_c = 0.1099$ where $\omega_c$ is the density parameter for cold
dark matter (CDM).  By varying $m_{3/2}$ or $f_{3/2}$, we can see the
effects of light gravitino on structure formation and CMB
anisotropies.  Moreover, we assume that neutrinos are massless in the
most part of this paper. We will make some comments on the
case where massive neutrinos are also included in Section~\ref{sec:summary}.

As briefly discussed in the introduction, the effects of WDM on
structure formation can be understood by considering following two
main aspects: (i) the change of the energy contents of the universe, or
the epoch of matter-radiation equality (unperturbed background
evolutions), (ii) the erasure of perturbations on small scales via
free-streaming (perturbation evolutions).  The first point is due to
the fact that WDM behaves as relativistic component at early times but
non-relativistic one at late times. Thus it changes the time of
matter-radiation equality depending on the mass.  It alters the
evolution of gravitational potential and drives the integrated
Sachs-Wolfe (ISW) effect in the CMB temperature anisotropy.  However,
in the case of light gravitino in which we are interested, its
abundance is so small $N_{3/2} = 0.059$ that the epoch of
radiation-matter equality is scarcely affected, even when we compare
the two opposite limits, $f_{3/2}=0$ ($m_{3/2}= 0$~eV) and $f_{3/2}=1$
($m_{3/2}= 86$~eV), with fixed $\omega_{\rm dm}=0.1099$.  Therefore it
is almost impossible to constrain the gravitino mass from unlensed CMB
anisotropies, even when the ideal observations (cosmic variance
limeted survey) are available.

\begin{figure}[htb]
\begin{center}
  \scalebox{1.0}{\includegraphics{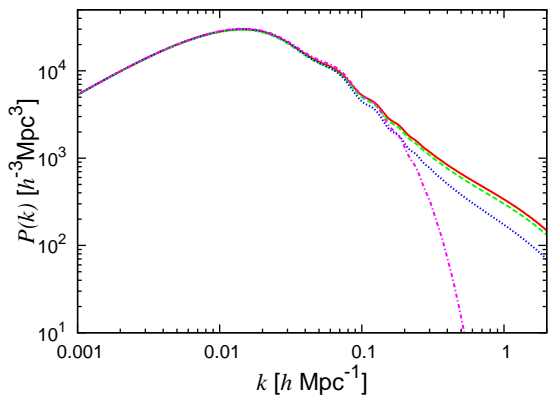}}
  \caption{Matter power spectra $P(k)$ for  several values of $m_{3/2}$.  We
    have plotted the cases with $m_{3/2}=0$ eV (solid red line), $1$
    eV (dashed green line), $10$ eV (dotted blue line), $86$ eV
    (dot-dashed magenta line).  In all cases we fixed the total dark
    matter density and energy density of gravitino as $\omega_{\rm
      dm}=0.1099$ and $N_{3/2}=0.059$, respectively.  
      We adopt HALOFIT\cite{Smith:2002dz} in calculating nonlinear corrections. }
\label{fig:mpk}
\end{center}
\end{figure}

Possible constraints on the gravitino mass almost come from the second
point.  Light gravitino free-streams to erase cosmic density
fluctuations while it is relativistic.  On scales smaller than the
free streaming length of gravitino, fluctuations of matter and hence
gravitational potential are erased.  To see the effect, we show
matter power spectra $P(k)$ for several values of $m_{3/2}$ with
$\omega_{\rm dm}$ being fixed in Fig.~\ref{fig:mpk}. We take
$m_{3/2}=0$ eV (solid red), $m_{3/2}=1$ eV (dashed green),
$m_{3/2}=10$ eV (dotted blue) and $m_{3/2}=86$ eV (dot-dashed
magenta).  As seen from the figure, as $m_{3/2}$ increases, the
suppression of $P(k)$ becomes larger on small scales, while on large
scales the amplitude of $P(k)$ is unaffected regardless of the value
of $m_{3/2}$.  With more careful observation we can notice that when
$m_{3/2}$ is small, the suppression of the power is small, but the
scale under which $P(k)$ is suppressed becomes large.  On the other
hand, when $m_{3/2}$ is large, the suppression is also large, however
the free-streaming scale becomes small.  These can be simply
understood as follows.  When the mass of gravitino is small,  gravitino
can erase cosmic structure up to large scales.  However, the smallness
of the mass in turn indicates that gravitino is minor component in the
contents of energy density and gravitationally irrelevant.  Thus
density fluctuations are less suppressed.  For the case of larger mass, 
the opposite argument holds.

\begin{figure}[htb]
\begin{center}
  \begin{tabular}{cc}
    \scalebox{1.0}{\includegraphics{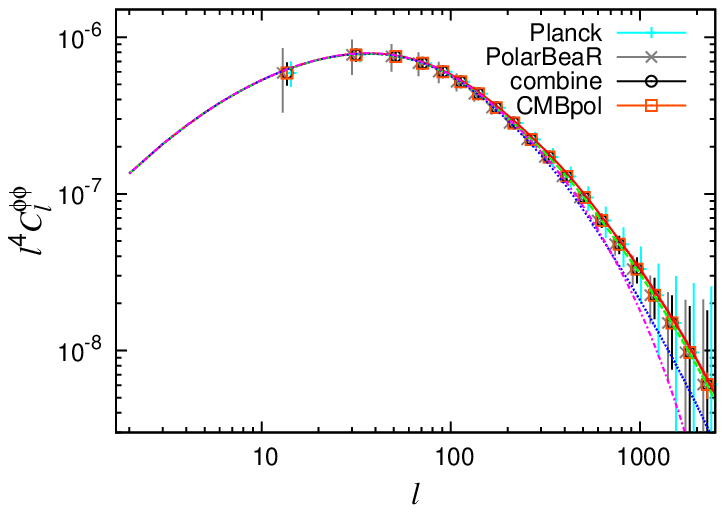}} &
    \scalebox{1.0}{\includegraphics{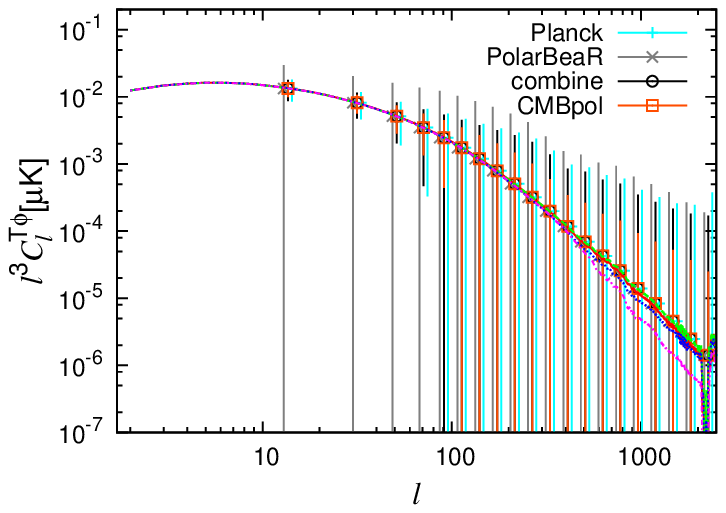}}
  \end{tabular}
  \caption{The angular power spectra of lensing potential
    $C^{\phi\phi}_\ell$ (Left) and its correlation with temperature
    $C^{T\phi}_\ell$ (Right).  The values of $m_{3/2}$ are same as in
    Figure~\ref{fig:mpk}. The sensitivities of future CMB surveys used in our
    analysis, Planck, PolarBeaR and CMBpol, 
    are also shown as points with error bars. 
    Notice that the bins are linearly and logarithmically spaced 
    in $\ell$ for $\ell\le100$ and $\ell>100$, respectively.
    For visual reason, data points for Planck and PolarBeaR are slightly displaced horizontally.}
\label{fig:phi}
\end{center}
\end{figure}

Now we move on to discuss how the suppression of matter fluctuations
changes the lensed CMB anisotropy.  CMB photons last-scattered at the
decoupling epoch, while traveling to the present epoch, are deflected
by the gravitational potential $\Phi(\mathbf{r}, \eta)$ generated by
the matter fluctuations (For a recent review see
e.g. \cite{Lewis:2006fu}).  The lensing potential $\phi$ is given by
\begin{equation}
\phi(\hat n) = -2\int^{\chi_*}_0 d\chi
\frac{\chi_*-\chi}{\chi_* \chi} \Phi(\chi \hat n, \eta(\chi)), 
\end{equation}
where $\chi$ is the comoving distance along the line of sight,
$\chi_*$ is the comoving distance to the last scattering surface and
$\eta(\chi)$ is the conformal time corresponds to the comoving
distance of $\chi$.  Actually, the lensing potential is not a direct
observable in CMB observations, but we can reconstruct it with
observed lensed CMB anisotropies. 
It contains much more
information than the lensed power spectrum~\cite{Smith:2006nk,Smith:2008an},
since the reconstruction is performed by making use of off-diagonal components in 
correlation function of lensed anisotropies~\cite{Okamoto:2003zw}.
In Fig.~\ref{fig:phi}, we show the angular power spectra of the lensing
potential and its correlation with the temperature anisotropy,
$C^{\phi\phi}_\ell$ and $C^{T\phi}_\ell$, respectively. For
  reference, we also show expected data of future CMB experiments:
  Planck, PolarBeaR and CMBpol. We can see from the
$C^{\phi\phi}_\ell$ in the Fig.~\ref{fig:phi} that the suppression of
the power spectra depends on the mass of gravitino, which can be
probed with future observations of CMB.  When we carefully observe the
power on small scales, the trend how the power is suppressed is
similar to what we have seen in the matter power spectra.  
Therefore we can
expect the mass of gravitino is constrained with reconstructed lensing
potential, which would be obtained from future CMB surveys.
Although cross correlation of the lensing potential with CMB temperature
anisotropy $C^{T\phi}_\ell$ is affected by the mass of gravitino, 
the effect is much small compared with the expected errors 
for future CMB surveys, which can be seen from the right panel in Fig.~\ref{fig:phi}.
Thus it is suggested that $C^{T\phi}_\ell$ has little advantage for 
constraining the mass of light gravitino.

Here it should be noted that some careful consideration must be given
for the following fact: the heavier the light gravitino mass is, the
smaller the free-streaming scale would be.  Although, regarding the suppression
of the power, the effects of the light gravitino is more significant
for a larger mass, when a survey cannot observe up to high multipoles
due to its limitation of the resolution, gravitino with lighter mass
can be better probed.  This is because gravitino with lighter mass can
erase cosmic structure up to larger scales compared to the case with
larger mass although the power suppression is milder.  As a simple
example, let us compare the two cases, $f_{3/2}=0$ and $f_{3/2}=1$
while keeping $\omega_{\rm dm}$ fixed.  At small angular scales, the
lensing potential $\phi$ for $f_{3/2}=1$ is more suppressed than that
for $f_{3/2}=0$.  However, 
for the case with $f_{3/2} = 1$ corresponding to $m_{3/2}\simeq 86$~eV, the free
streaming length is small. Therefore the suppression occurs only at
limited small angular scales.  If the observed multipoles are limited
to low $\ell$'s, where the suppression cannot be seen, gravitino with
$f_{3/2}=1$ cannot leave any imprint on such a measurement, which
means that we cannot differentiate models between $f_{3/2}=1$ and
$f_{3/2}=0$.  This makes the likelihood surface multi-modal and
highly-degenerate.  To break up these degeneracies, high-resolution
measurement of lensing potential is needed, and currently available
observations cannot suffice this requirement.  In the next section, we
discuss how future CMB surveys will constrain the light gravitino
models.

\section{Constraints on light gravitino mass}\label{sec:constraints}
Now in this section, we investigate the constraints on the light gravitino
mass from future CMB surveys.  As discussed in the previous
section, since the current CMB surveys are not precise enough to 
measure the lensing potential, it is almost impossible 
to probe $m_{3/2}$. However, in future surveys of CMB, 
the measurement of lensing potential would be significantly improved. 
To forecast constraints on light gravitino mass from
future CMB surveys, we make use of the following three surveys in this
paper, the Planck~~\cite{:2006uk}, PolarBeaR~\cite{PolarBeaR} and
CMBpol~\cite{CMBpol}.  The parameters for instrumental
design for these surveys are summarized in Table~\ref{table:surveys}, where
$\theta_{FWHM}$ is the size of Gaussian beam\footnote{
  We assume Gaussian beam and neglect any anisotropies in beam and
  distortion arising from the scan strategy.
} 
at FWHM and $\sigma_T$ ($\sigma_P$) is the temperature
(polarization) noise.

\begin{table}[htb]
  \begin{center}
  \begin{tabular}{l||c|c|c|c|c}
  \hline
  \hline
  surveys & $f_{\rm sky}$ & bands [GHz] & $\theta_{\rm FWHM}$ [arcmin] & $\sigma_T$ [$\mu$K] & $\sigma_P$ [$\mu$K] \\
  \hline
  Planck~\cite{:2006uk}
  & $0.65$ & $100$ & $9.5$ & $6.8$ & $10.9$\\
  && $143$ & $7.1$ & $6.0$ & $11.4$\\
  && $217$ & $5.0$ & $13.1$ & $26.7$\\
  \hline
  PolarBeaR~\cite{PolarBeaR}
  & $0.03$ & $90$ & $6.7$ & $1.13$ & $1.6$\\
  && $150$ & $4.0$ & $1.70$ & $2.4$\\
  && $220$ & $2.7$ & $8.00$ & $11.3$\\
  \hline
  CMBpol~\cite{CMBpol} 
  & $0.65$ & $100$ & $4.2$ & $0.84$ & $1.18$\\
  && $150$ & $2.8$ & $1.26$ & $1.76$\\
  && $220$ & $1.9$ & $1.84$ & $2.60$\\
  \hline
  \hline 
\end{tabular}
\caption{Instrumental parameters for future CMB surveys used in our
  analysis.  $\theta_{\rm FWHM}$ is Gaussian beam width at FWHM,
  $\sigma_T$ and $\sigma_P$ are temperature and polarization noise,
  respectively.  For the Planck and PolarBeaR surveys, we assume
  1-year duration of observation and for the CMBpol survey, we assumed
  4-year duration.  }
  \label{table:surveys}
\end{center}
\end{table}

In this paper, to generate samples from the Bayesian posterior distributions of
cosmological parameters, we make use of the public code {\tt
  MultiNest} \cite{Feroz:2007kg} integrated in the vastly used 
 Monte Carlo sampling code {\tt
  COSMOMC}~\cite{Lewis:2002ah}.  While {\tt COSMOMC} samples the
posterior distributions via the Markov chain Monte Carlo (MCMC) 
sampling method, {\tt MultiNest} is based on the different sampling called 
nested-sampling method \cite{Skilling:2004}. 
Use of {\tt MultiNest} has several advantages in our analysis. One of
the greatest advantages is that it enables efficient exploring of
multi-modal/highly-degenerate likelihood surface, which is indeed the
case for light gravitino models, as we have discussed in the previous
section.  Also it provides Bayesian evidence of a model and hence
enables us to employ Bayesian model selection.

\begin{table}[htb]
  \begin{center}
  \begin{tabular}{l||c|c|c}
  \hline
  \hline
  & & \multicolumn{2}{c}{prior ranges} \\
  \cline{3-4}
  parameters & fiducial values & Planck/PolarBeaR/combined & CMBpol\\
  \hline
  $\omega_b$ & $0.02273$ & $0.018 \to 0.28$ & $0.021 \to 0.024$\\
  $\omega_{\rm dm}$ & $0.1099$ & $0.08 \to 0.30$ & $0.10 \to 0.14$\\
  $\theta_s$ & $1.0377$ & $1.02 \to1.06$ & $1.03 \to 1.04$\\
  $\tau$ & $0.087$ & $0.01 \to 0.30$ & $0.06 \to 0.14$\\
  $N_{3/2}$ & $0.059$ & ($0 \to 5$) & ($0 \to 2$)\\
  $f_{3/2}$ & $0.013$ & $0 \to 1$ & $0 \to 0.1$\\
  $Y_p$ & $0.248$& $(0.1 \to 0.5)$ & ($0.2 \to 0.3$)\\
  $n_s$ & $0.963$ & $0.8 \to 1.2$ & $0.9 \to 1$\\
  $\ln(10^{10}A_s)$ & $3.063$ & $2.8\to3.5$ & $3.0 \to 3.2$\\
  \hline
  \hline 
\end{tabular}
\caption{The fiducial values and prior ranges for the parameters used
  in the analysis.  Note that priors shown with parenthesis are
  imposed only when the corresponding parameters ($N_{3/2}$ and $Y_p$)
  are treated as free parameters and not imposed when they are fixed
  or derived from other parameters. For CMBpol, we take narrower range
  for the top priors since its accuracy is much higher than the former
  two surveys.  Hence we do not need broad range for the priors.}
  \label{table:priors}
\end{center}
\end{table}

To obtain the limit for the mass of light gravitino, we can translate
the constraint on the parameter $f_{3/2}= \omega_{3/2} / \omega_{\rm
  dm}$ using Eq.~\eqref{eq:omega32}.  Since light gravitino has almost
definite prediction of its abundance, we mainly report our results for
the case with $N_{3/2}=0.059$ being fixed. However, in some scenario
such as those with late-time entropy production, this number may be
altered.  In this respect, we also make analysis with $N_{3/2}$ being
varied.  Furthermore, regarding the treatment of the primordial
abundance of $^4$He (denoted as $Y_p$), we assume two cases: treating $Y_p$ as a
free parameter and fixing $Y_p$ with the derived value via the
big bang nucleosynthesis (BBN) relation~\cite{Kawanocode}.  
In the BBN theory, $Y_p$ is
determined once baryon density $\omega_{\rm b}$ and the effective
number of neutrino $N_{\rm eff}$ are given.  Thus such a fixing of the 
value of $Y_p$ was adopted in some analysis \cite{Ichikawa:2006dt,Hamann:2007sb,Ichikawa:2007js}.
Since, in the precise
measurement of future CMB survey, the prior on $Y_p$ can also affect
the determination of other cosmological parameters
\cite{Hamann:2007sb,Ichikawa:2007js}, we consider the case with $Y_p$ freely
varied as well.

Thus the full parameter space that we explore for light gravitino
models are basically nine-dimensional:
\begin{equation}
(\omega_b, \omega_{\rm dm}, \theta_s, \tau, N_{3/2}, f_{3/2}, Y_p, n_s, A_s),
\end{equation}
where $\theta_s$ is the acoustic peak scale, $\tau$ is the optical
depth of reionization and $A_s$ and $n_s$ are the amplitude and
spectral index of initial power spectrum of scalar perturbations at a
pivot scale $k_0 = 0.05$ Mpc$^{-1}$.  In the following, we investigate
four different cases: (I) fixing $N_{3/2}=0.059$ and deriving $Y_p$ via
the BBN relation, (II) fixing
$N_{3/2}=0.059$ and treating $Y_p$ as a free parameter, (III) treating $N_{3/2}$ as a free parameter and
$Y_p$ as a derived parameter via the BBN relation, and (IV)
treating $N_{3/2}$ and $Y_p$ as free parameters.  The fiducial values
and top-hat priors for parameter estimation are summarized in
Table~\ref{table:priors}.

The likelihood function is adopted from Ref.~\cite{Perotto:2006rj}.
We include $TT, TE, EE, \phi\phi, T\phi$ spectra for correlation of
CMB anisotropy and lensing potential up to $\ell \le 2500$. 
We assume 
lensing reconstruction is performed 
by adopting the method based on quadratic estimator \cite{Okamoto:2003zw}, 
and the expected noise in lensing potential is
calculated by the publicly available {\tt FuturCMB2} code developed by the
authors of \cite{Perotto:2006rj}.
Angular power spectra are calculated
using the method in Ref.~\cite{Challinor:2005jy}.  For corrections for
lensing potential due to nonlinear evolution of matter density
perturbations, we adopt HALOFIT, which is based on the N-body
simulations of CDM models \cite{Smith:2002dz}.  Though the light gravitino model is not
exactly the CDM models, we believe that the change of the nonlinear correction
is negligible.  This is because that gravitino has small $N_{3/2}$ and
regardless of the mass of $m_{3/2}$, dark matter can be approximated
by CDM very well when it begins to evolve in nonlinear regime.
Furthermore, nonlinear correction changes the spectra of lensing
potential by only a few percent at $\ell \le
2500$~\cite{Challinor:2005jy}, and hence the treatment here can be
justified. 
In addition, we also performed same analyses without 
including nonlinear corrections and 
checked that resultant constraints do not significantly change 
by the treatment of nonlinearity.

\begin{figure}[htb]
\begin{center}
  \begin{tabular}{ccc}
    \scalebox{1.0}{\includegraphics{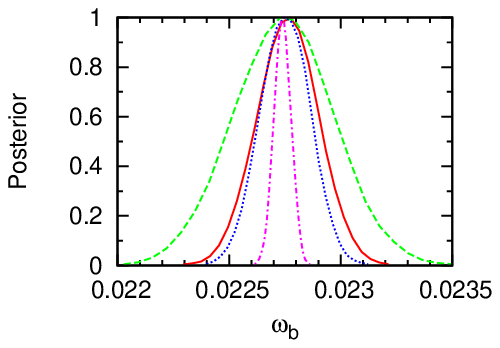}} &
    \scalebox{1.0}{\includegraphics{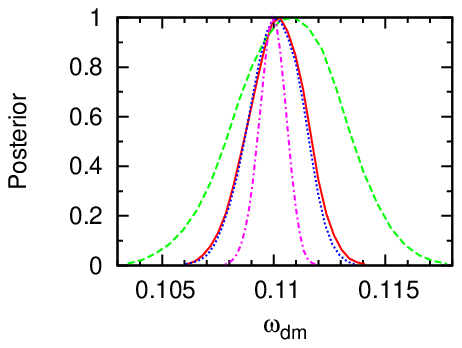}} &
    \scalebox{1.0}{\includegraphics{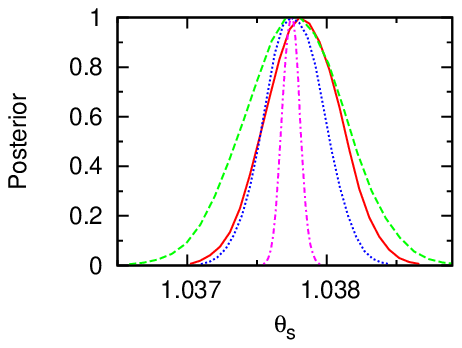}} \\
    \scalebox{1.0}{\includegraphics{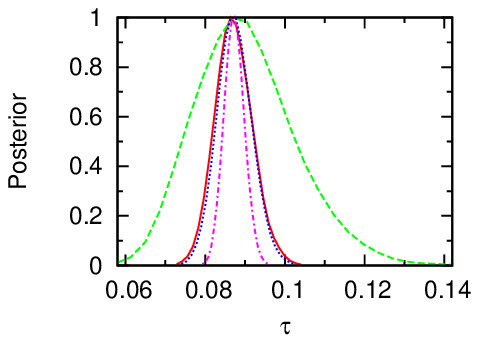}} &
    \scalebox{1.0}{\includegraphics{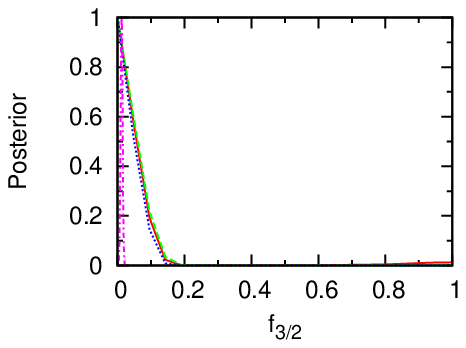}} &
    \scalebox{1.0}{\includegraphics{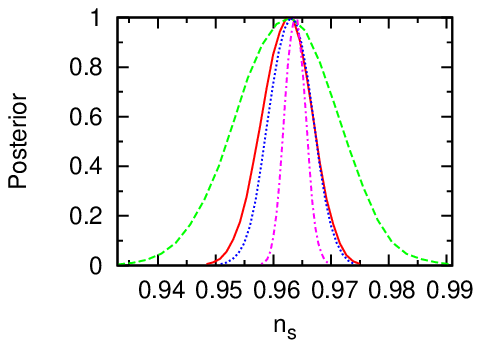}} \\
    \scalebox{1.0}{\includegraphics{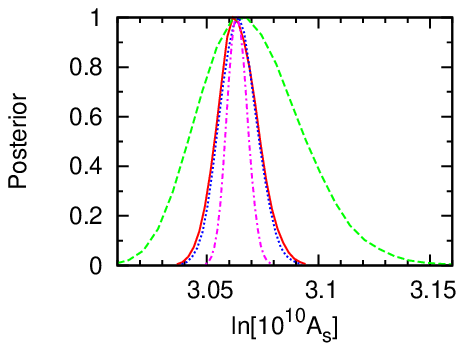}} & & 
  \end{tabular}
  \caption{The 1d posterior distributions for cosmological parameters
    with fixed $N_{3/2}=0.059$ and the BBN relation for
    $Y_p$ adopted.  Shown are constraints from Planck alone (red solid
    line), PolarBeaR alone (green dashed line), Planck and PolarBeaR
    combined (blue dotted line), and CMBpol (magenta dash-dotted line).
  }
\label{fig:1d_fixed}
\end{center}
\end{figure}

Now we are going to present our results.  In
Tables~\ref{table:case1}-\ref{table:case4} we summarize the
constraints on the cosmological parameters
from Planck alone, PolarBeaR alone, Planck and PolarBeaR combined, 
and CMBpol alone, separately for different priors. 
First we discuss the case
with fixed $N_{3/2}=0.059$ and the BBN relation adopted for $Y_p$.
The 1d posterior distributions of cosmological parameters are shown in
Fig.~\ref{fig:1d_fixed}.  From the posterior distributions for
$f_{3/2}$ in Fig.~\ref{fig:1d_fixed}, we can easily see that light
gravitino models are not constrained very well with Planck or
PolarBeaR alone.  The posterior distributions have decaying
tails from the peak near $f_{3/2}=0$ to $f_{3/2}=1$.  Actually, they
have very smooth second peaks at around $f_{3/2}=1$.  This multi-modal
structure of posterior distributions comes from the degeneracies what
we have discussed in Section~\ref{sec:CMB}.  Light gravitino with a
relatively large mass ($f_{3/2}\simeq 1$) can suppress the power via
free-streaming only at very small scales where the Planck surveys
cannot sufficiently measure the CMB.  Thus a model with such a
gravitino mass can fit the data from Planck alone.  On the other hand,
PolarBeaR has better resolution, gravitino with large mass is much
constrained from observation of lensing potential at small scales.
However, the sky coverage of the PolarBeaR survey is much smaller than
Planck, thus the observation is worse at large angular scales.  In
this case, other parameters than $f_{3/2}$ are still not
well-constrained from PolarBeaR alone.  Therefore gravitino with
relatively large mass can fit the data to some extent by tuning other
cosmological parameters in this case too.  To remove the degeneracy it
is necessary to combine observations precise at large and small
angular scales or, ultimately, use measurements precise both at large
and small angular scales.  With data from Planck and PolarBeaR
combined, we can obtain a constraint
\begin{equation}
f_{3/2}\le0.036 \ \mbox{(95\% C.L.)}.
\end{equation}
When we use CMBpol, whose measurement is precise both 
on large and small scales than other two survey, 
this constraint can be improved as
\begin{equation}
f_{3/2}=0.0121\pm0.0027 \ \mbox{(68\% C.L.)}.
\end{equation}
These constraints are translated into the limits on the mass of light
gravitino.  For the case with Planck and PolarBeaR combined,
the constraint is given as
\begin{equation}
m_{3/2}\le3.2 \ \mbox{eV} \ \mbox{(95\% C.L.)}, \label{eq:combined}
\end{equation}
and with CMBpol as 
\begin{equation}
m_{3/2}=1.04^{+0.22}_{-0.26} \ \mbox{eV} \ \mbox{(68\% C.L.)}.
\end{equation}
Since gravitino mass should be larger than $1$eV not to destabilize
the messenger scalar, CMBpol would be expected to give
(counter-)evidense for existence of gravitino if its mass is (not) in
the mass range considered here.  In Section~\ref{sec:selection} we will
discuss this point more quantitatively using Bayesian models selection
analysis.

\begin{figure}[htb]
\begin{center}
  \begin{tabular}{cc}
    \scalebox{1.3}{\includegraphics{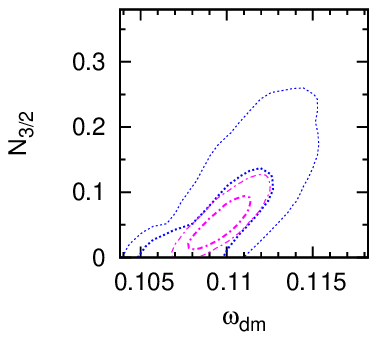}} & \\
    \scalebox{1.3}{\includegraphics{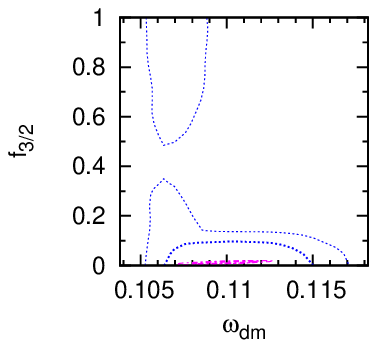}} &
    \scalebox{1.3}{\includegraphics{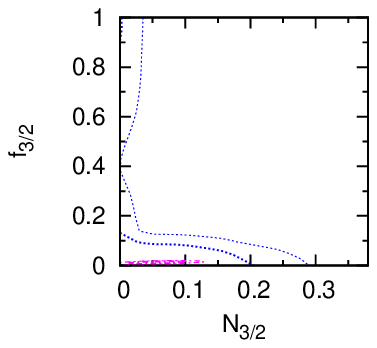}} 
  \end{tabular}
  \caption{The degeneracies arise among $\omega_{\rm dm}$, $f_{3/2}$ and $N_{3/2}$.
  Shown are the cases where we adopted the BBN relation for $Y_p$ using 
  data from Planck and PolarBeaR combined (blue dotted line) and 
  CMBpol alone (magenta dash-dotted line). 
  The thick and thin lines show contours at 68 \% and 95\% C.L., respectively}
\label{fig:2d_deg}
\end{center}
\end{figure}

\begin{figure}[htb]
\begin{center}
  \begin{tabular}{ccc}
    \scalebox{1.0}{\includegraphics{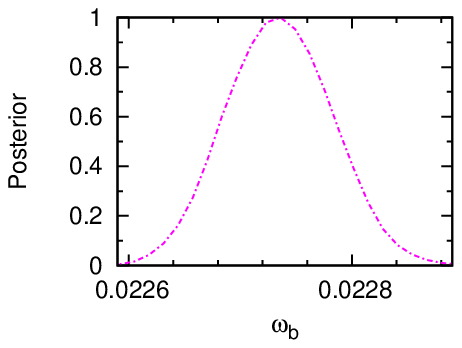}} &
    \scalebox{1.0}{\includegraphics{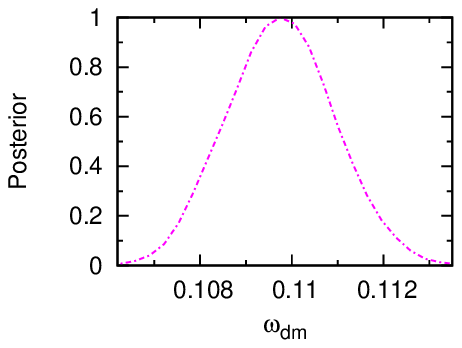}} &
    \scalebox{1.0}{\includegraphics{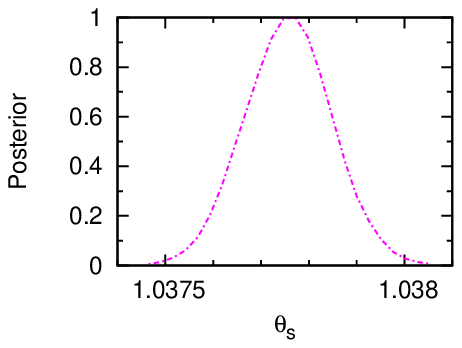}} \\
    \scalebox{1.0}{\includegraphics{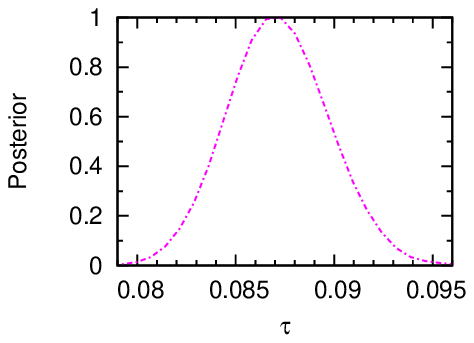}} &
    \scalebox{1.0}{\includegraphics{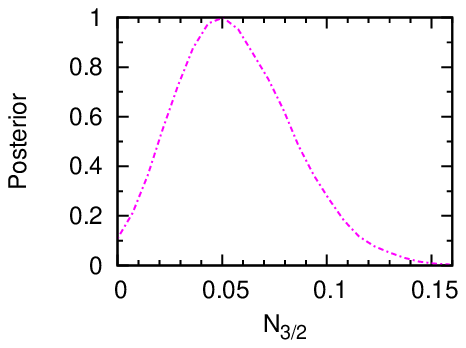}} &
    \scalebox{1.0}{\includegraphics{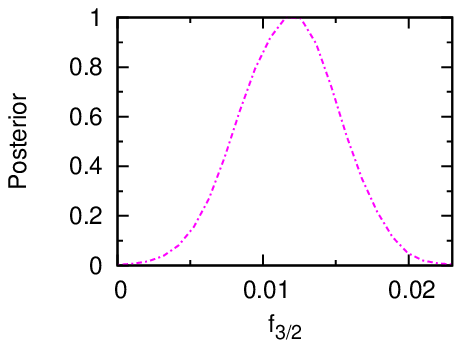}} \\
    \scalebox{1.0}{\includegraphics{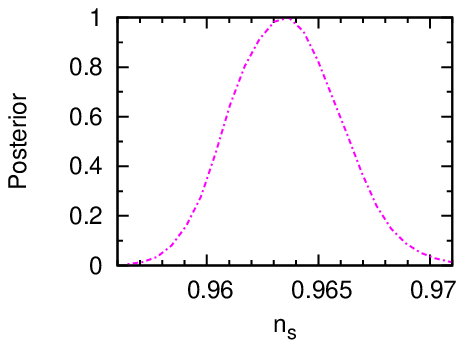}} &
    \scalebox{1.0}{\includegraphics{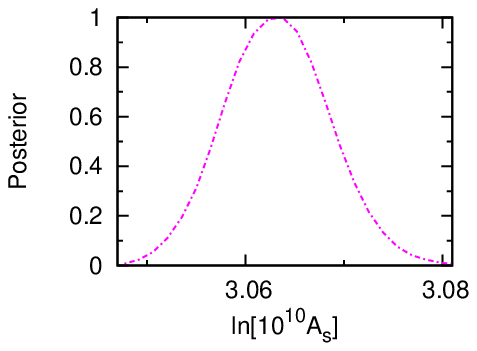}} & 
  \end{tabular}
  \caption{The 1d posterior distributions for cosmological parameters
    with freely varying  $N_{3/2}$ and the BBN relation for $Y_p$ adopted. 
    The case for CMBpol data is only shown.  }
\label{fig:1d_free}
\end{center}
\end{figure}

So far we kept assuming the energy density of gravitino fixed as
$N_{3/2}=0.059$.  If we loosen this assumption and take $N_{3/2}$ as a
free parameter, the constraints are significantly weakened.  
Notice that the free-streaming scale of light gravitino
is determined by its mass. 
Thus when $N_{3/2}$ is freely varied and hence the mass of gravitino can be large,  
the free-streaming scale can be shifted toward 
smaller scales  over which PolarBeaR cannot observe for a wide range of the mass
and one cannot see the damping 
of the power there. (On the other hand, since CMBpol is very precise on small scales,
the constraint on $m_{3/2}$ from CMBpol is not affected much.)
Thus we cannot obtain a meaningful constraint on $f_{3/2}$ even if we 
combine Planck and PolarBeaR when we take $N_{3/2}$ as a free parameter.
Furthermore, the change in $N_{3/2}$ renders the shift of the radiation-matter equality
which can be absorbed by tuning $\omega_{\rm dm}$ 
\cite{Hannestad:2003ye, Crotty:2004gm, Ichikawa:2008pz}, 
significant degeneracies arise among 
$\omega_{\rm dm}$, $f_{3/2}$ and $N_{3/2}$ as shown in 
Fig.~\ref{fig:2d_deg}\footnote{
  In Figure 7 and 12 in~\cite{Boyarsky:2008xj}, 
  a similar degeneracy can also be seen as band-like 
  allowed region along $F_\mathrm{WDM}$ axis.
}.
For meaningful
constraints we need sensitivities as good as those of CMBpol-like
survey. 
 In Fig.~\ref{fig:1d_free} 1d posterior distributions for
parameters including $N_{3/2}$ are shown. The 68 \% limit of $f_{3/2}$
for this case is $f_{3/2}=0.0118^{+0.0032}_{-0.0031}$, which
corresponds to the constraint on the gravitino mass
$m_{3/2}=1.19^{+0.16}_{-0.50}$~eV.

\begin{table}[htb]
  \begin{center}
  \begin{tabular}{l||cccc}
  \hline
  \hline
  parameters & Planck & PolarBeaR & combined & CMBpol\\
  \hline
  $100~\omega_b$ & $2.276^{+0.013}_{-0.012}$ & $2.274^{+0.021}_{-0.022}$ & $2.275^{+0.012}_{-0.009}$ & $2.2739^{+0.0041}_{-0.0033}$ \\
  $\omega_{dm}$ & $0.1101^{+0.0013}_{-0.0010}$ & $0.1106^{+0.0022}_{-0.0022}$ & $0.1100^{+0.0012}_{-0.0010}$ & $0.10993^{+0.00058}_{-0.00059}$ \\
  $100~\theta_s$ & $103.783^{+0.024}_{-0.026}$ & $103.777^{+0.035}_{-0.031}$ & $103.777^{+0.020}_{-0.019}$ & $103.774^{+0.0056}_{-0.0061}$ \\
  $\tau$ & $0.0871^{+0.0043}_{-0.0045}$ & $0.090^{+0.009}_{-0.013}$ & $0.0873^{+0.0039}_{-0.0043}$ & $0.0872^{+0.0022}_{-0.0026}$ \\
  $N_{3/2}$ & --- & --- & --- & --- \\
  $f_{3/2}$ & --- & --- & $<0.036$ (95\%) & $0.0121^{+0.0027}_{-0.0027}$ \\
  $Y_p$ & --- & --- & --- & --- \\
  $n_s$ & $0.9622^{+0.0045}_{-0.0036}$ & $0.9621^{+0.0081}_{-0.0083}$ & $0.9629^{+0.0036}_{-0.0033}$ & $0.9637^{+0.0017}_{-0.0017}$ \\
  $\ln(10^{10}\times A_s)$ & $3.0640^{+0.0073}_{-0.0094}$ & $3.077^{+0.015}_{-0.026}$ & $3.0641^{+0.0063}_{-0.0094}$ & $3.0637^{+0.0040}_{-0.0047}$ \\
  \hline
  $m_{3/2}$~[eV] & --- & --- & $<3.2$ (95\%) & $1.04^{+0.22}_{-0.26}$ \\
  \hline  
  \hline \end{tabular}
  \caption{Constraints on cosmological parameters for the case with fixing
  $N_{3/2}=0.059$ and adopting the BBN relation to fix the value  of $Y_p$ (CASE I). 
  We basically present the mean values as well as $1\sigma$ errors. 
   For parameters that are bounded only 
  from one side we present 95\% credible intervals.}
  \label{table:case1}
\end{center}
\end{table}

\begin{table}[htb]
  \begin{center}
  \begin{tabular}{l||cccc}
  \hline
  \hline
  parameters & Planck & PolarBeaR & combined & CMBpol\\
  \hline
  $100~\omega_b$ & $2.278^{+0.021}_{-0.017}$ & $2.272^{+0.035}_{-0.034}$ & $2.274^{+0.016}_{-0.014}$ & $2.2722^{+0.0055}_{-0.0047}$ \\
  $\omega_{dm}$ & $0.1100^{+0.0015}_{-0.0010}$ & $0.1107^{+0.0025}_{-0.0021}$ & $0.1100^{+0.0011}_{-0.0011}$ & $0.10990^{+0.00061}_{-0.00056}$ \\
  $100~\theta_s$ & $103.788^{+0.048}_{-0.042}$ & $103.775^{+0.067}_{-0.058}$ & $103.775^{+0.032}_{-0.035}$ & $103.771^{+0.011}_{-0.010}$ \\
  $\tau$ & $0.0873^{+0.0041}_{-0.0050}$ & $0.090^{+0.011}_{-0.013}$ & $0.0873^{+0.0040}_{-0.0044}$ & $0.0870^{+0.0020}_{-0.0028}$ \\
  $N_{3/2}$ & --- & --- & --- & --- \\
  $f_{3/2}$ & --- & --- & $<0.035$ (95\%) & $0.0118^{+0.0030}_{-0.0025}$ \\
  $Y_p$ & $0.250^{+0.008}_{-0.011}$ & $0.248^{+0.016}_{-0.016}$ & $0.2486^{+0.0094}_{-0.0066}$ & $0.2481^{+0.0031}_{-0.0028}$ \\
  $n_s$ & $0.9627^{+0.0067}_{-0.0063}$ & $0.961^{+0.012}_{-0.014}$ & $0.9625^{+0.0063}_{-0.0052}$ & $0.9629^{+0.0026}_{-0.0025}$ \\
  $\ln(10^{10}\times A_s)$ & $3.064^{+0.008}_{-0.010}$ & $3.071^{+0.018}_{-0.028}$ & $3.064^{+0.006}_{-0.010}$ & $3.0631^{+0.0043}_{-0.0045}$ \\
  \hline
  $m_{3/2}$~[eV] & --- & --- & $<3.1$ (95\%) & $1.02^{+0.26}_{-0.22}$ \\
  \hline
  \hline 
  \end{tabular}
  \caption{Same tables as in Table~\ref{table:case1} but for the cases with fixing $N_{3/2}=0.059$ and treating $Y_p$ as a free parameter (CASE II).}
  \label{table:case2}
\end{center}
\end{table}

\begin{table}[htb]
  \begin{center}
  \begin{tabular}{l||cccc}
  \hline
  \hline
  parameters & Planck & PolarBeaR & combined & CMBpol\\
  \hline
  $100~\omega_b$ & $2.276^{+0.015}_{-0.015}$ & $2.284^{+0.023}_{-0.030}$ & $2.276^{+0.011}_{-0.014}$ & $2.2735^{+0.0043}_{-0.0048}$ \\
  $\omega_{dm}$ & $0.1099^{+0.0022}_{-0.0034}$ & $0.1125^{+0.0036}_{-0.0060}$ & $0.1097^{+0.0017}_{-0.0033}$ & $0.1097^{+0.0011}_{-0.0011}$ \\
  $100~\theta_s$ & $103.782^{+0.035}_{-0.029}$ & $103.763^{+0.053}_{-0.038}$ & $103.778^{+0.027}_{-0.027}$ & $103.776^{+0.0092}_{-0.0082}$ \\
  $\tau$ & $0.0873^{+0.0040}_{-0.0047}$ & $0.091^{+0.008}_{-0.015}$ & $0.0874^{+0.0036}_{-0.0046}$ & $0.0871^{+0.0022}_{-0.0027}$ \\
  $N_{3/2}$ & $<0.24$ (95\%) & $<0.47$ (95\%) & $<0.21$ (95\%) & $<0.10$ (95\%) \\
  $f_{3/2}$ & --- & --- & --- & $0.0118^{+0.0032}_{-0.0031}$ \\
  $Y_p$ & --- & --- & --- & --- \\
  $n_s$ & $0.9638^{+0.0034}_{-0.0065}$ & $0.968^{+0.009}_{-0.012}$ & $0.9639^{+0.0038}_{-0.0054}$ & $0.9635^{+0.0020}_{-0.0025}$ \\
  $\ln(10^{10}\times A_s)$ & $3.064^{+0.009}_{-0.010}$ & $3.076^{+0.021}_{-0.027}$ & $3.064^{+0.008}_{-0.010}$ & $3.0632^{+0.0051}_{-0.0049}$ \\
  \hline 
  $m_{3/2}$~[eV] & --- & --- & --- & $1.19^{+0.16}_{-0.50}$ \\
  \hline
  \hline \end{tabular}
  \caption{Same tables as in Table~\ref{table:case1} but for the cases with treating $N_{3/2}$ as a free parameter and adopting the BBN relation (CASE III).}
  \label{table:case3}
\end{center}
\end{table}

\begin{table}[htb]
  \begin{center}
  \begin{tabular}{l||cccc}
  \hline
  \hline
  parameters & Planck & PolarBeaR & combined & CMBpol\\
  \hline
  $100~\omega_b$ & $2.278^{+0.017}_{-0.021}$ & $2.276^{+0.030}_{-0.035}$ & $2.273^{+0.012}_{-0.013}$ & $2.2727^{+0.0046}_{-0.0056}$ \\
  $\omega_{dm}$ & $0.1117^{+0.0023}_{-0.0056}$ & $0.1176^{+0.0062}_{-0.0092}$ & $0.1103^{+0.0017}_{-0.0041}$ & $0.1103^{+0.0013}_{-0.0016}$ \\
  $100~\theta_s$ & $103.75^{+0.10}_{-0.06}$ & $103.65^{+0.14}_{-0.11}$ & $103.760^{+0.068}_{-0.045}$ & $103.764^{+0.027}_{-0.019}$ \\
  $\tau$ & $0.0875^{+0.0042}_{-0.0049}$ & $0.091^{+0.008}_{-0.015}$ & $0.0872^{+0.0036}_{-0.0031}$ & $0.0872^{+0.0021}_{-0.0025}$ \\
  $N_{3/2}$ & $<0.47$ (95\%) & $<1.3$ (95\%) & $<0.34$ (95\%) & $<0.17$ (95\%) \\
  $f_{3/2}$ & --- & --- & --- & $0.0124^{+0.0032}_{-0.0034}$ \\
  $Y_p$ & $0.2458^{+0.015}_{-0.011}$ & $0.230^{+0.029}_{-0.018}$ & $0.246^{+0.010}_{-0.007}$ & $0.2471^{+0.0043}_{-0.0043}$ \\
  $n_s$ & $0.9643^{+0.0068}_{-0.0068}$ & $0.965^{+0.014}_{-0.013}$ & $0.9636^{+0.0047}_{-0.0045}$ & $0.9632^{+0.0024}_{-0.0027}$ \\
  $\ln(10^{10}\times A_s)$ & $3.067^{+0.009}_{-0.011}$ & $3.082^{+0.019}_{-0.032}$ & $3.0636^{+0.0079}_{-0.0077}$ & $3.0641^{+0.0046}_{-0.0053}$ \\
  \hline
  $m_{3/2}$~[eV] & --- & --- & --- & $1.10^{+0.07}_{-0.61}$ \\
  \hline 
  \hline 
\end{tabular}
  \caption{Same tables as in Table~\ref{table:case1} but for the cases with treating both $N_{3/2}$ and $Y_p$ as free parameters (CASE IV).}
  \label{table:case4}
\end{center}
\end{table}

\section{Model selection analysis on light gravitino model}\label{sec:selection}
In the previous section, we have seen that future CMB surveys
give rather tight constraints on mass of light gravitino, so that 
we can expect they would give (counter-)evidence for existence 
of gravitino to more or less extent.
But then a question arises how strong the evidence for gravitino is.
This is a kind of model selection problem in statistics theory,
which has been often argued in cosmology~
\cite{Slosar:2002dc,
Beltran:2005xd,Trotta:2005ar,Mukherjee:2005wg,Bridges:2005br,Kunz:2006mc,
Magueijo:2006we,Liddle:2006tc,Pahud:2007gi,Heavens:2007ka,Bevis:2007gh,Trotta:2008qt}.
In Bayesian statistics, the natural measure for evidence of a model is 
{\it Bayesian evidence} $E$, 
\begin{equation}
E(M)=\int d\theta P(data|\theta) \pi(\theta|M),
\end{equation}
where $\theta$ represents a set of parameters included in a model $M$. 
$P(data|\theta)$ and $\pi(\theta|M)$ are
the likelihood and prior probability functions, respectively. 
Bayesian evidence can be efficiently calculated by nested sampling method~\cite{Skilling:2004}.
Predictiveness of a model $M_1$ against another $M_2$ can be assessed
by differencing the logarithm 
of Bayes factors of the models, that is
\begin{equation}
B_{12}=\ln(E(M_1)/E(M_2)),
\end{equation}
which is called Bayes factor. 
If $B_{12}$ is positively (negatively) large, we can say the observed data can be explained
well by model $M_1$ ($M_2$) compared with $M_2$ ($M_1$). 
As a rule of thumb the Jeffreys' scale is often used to translate a Bayes factor into literal expression 
for strength of an evidence: $B_{12}<1$ is not significant, $1<B_{12}<2.5$ significant,
$2.5<B_{12}<5$ strong and $5<B_{12}$ is decisive.
For more details we refer to a recent review \cite{Trotta:2008qt} and references therein.

Now we are going to see how large the evidence is from future CMB surveys.
In Table~\ref{table:evidence} we summarized values of obtained Bayes factor 
for light gravitino model with different sets of data and priors against 
the conventional CDM model ($f_{3/2}=N_{3/2}=0$).
Here we have assumed a same fiducial model 
($m_{3/2}=1$eV and $N_{3/2}=0.059$), 
as in the previous section.

\begin{table}
  \begin{center}
  \begin{tabular}{lrrrr}
    \hline
    \hline
    datasets & CASE I & CASE II & CASE III & CASE IV \\
    \hline
    Planck alone & $-3.36\pm0.17$ & $-2.81\pm0.17$ & $-5.76\pm0.17$ & $-4.70\pm0.18$ \\
    PolarBeaR alone & $-3.39\pm0.15$ & $-2.66\pm0.18$ & $-5.02\pm0.16$ & $-3.62\pm0.18$ \\
    combined & $-3.03\pm0.17$ & $-2.86\pm0.18$ & $-5.80\pm0.17$ & $-5.82\pm0.18$ \\
    CMBpol alone & $3.40\pm0.16$ & $3.63\pm0.17$ &  $1.08\pm0.17$ & $1.44\pm0.19$ \\
    \hline
    \hline
  \end{tabular}
  \caption{Bayes factors for the light gravitino model against the CDM model. Shown are the mean and
  standard errors from two independent samplings.}
  \label{table:evidence}
  \end{center}
\end{table}

First of all, from Table~\ref{table:evidence} 
we can see that Planck or PolarBeaR alone 
and even Planck and PolarBeaR combined give only negative 
Bayes factor for the gravitino model against the CDM model, 
regardless of priors on $Y_p$ and $N_{3/2}$.
This is because for most of values of added parameters $f_{3/2}$ (and $N_{3/2}$) 
from the CDM model, gravitino model can only marginally improve fit to the data, 
even though it indeed improves the fit at some values around the fiducial ones, 
$f_{3/2}\simeq0.013$ (and $N_{3/2}\simeq0.059$). 
In other words, the complexity of the gravitino model has little advantage in explaining the data.

The situation dramatically changes 
for the case of the CMBpol survey.
From Table~\ref{table:evidence} 
we can see from CMBpol data we obtain Bayes factor 
for gravitino model against the CDM model as 
$\ln B=3.40\pm0.16$ for a case with fixing $N_{3/2}=0.059$ and using the BBN relation.
This is interpreted as strong evidence in the Jeffreys' scale, 
though there's always some disagreement in that 
how much Bayes factor can be regarded as giving enough evidence.
For the cases with treating $N_{3/2}$ as a free parameter and using the BBN relation, 
we obtain $\ln B\gtrsim1.08\pm 0.17$. This can be regarded as giving 
only marginal evidence. So it is difficult to obtain enough evidence for 
general WDM model whose number density is not theoretically limited in some small range.
Fortunately, since gravitino has small model-dependence of $N_{3/2}$ we can take the former
value of $\ln B$.

So far we have discussed the case of fiducial gravitino mass $m_{3/2}$. 
For larger gravitino mass, as long as it is less than the current bound 
($m_{3/2}\lesssim 16$ eV~\cite{Viel:2005qj,Boyarsky:2008xj}), 
the evidence surely improves.
This is because for this range of mass, the power spectra of lensing potential
differ more and more from those for the CDM model as the gravitino mass increases.
Since, in Section~\ref{sec:model} 
we have seen that gravitino mass is expected to be $O(1)$ eV or larger theoretically, 
the fiducial model of $m_{3/2}=1$ eV, which we used throughout this paper, 
can be supposed as a rather pessimistic case.
Since we have seen that even evidence for gravitino with mass 1 eV can be 
probed by a CMBpol-like survey, we would expect such a survey can probe 
most part of theoretically-motivated range 
of light gravitino mass. 
We hope such a survey would be realized and probe 
light gravitino model in the near future.

\bigskip

\section{Summary and discussion}\label{sec:summary}

We investigated a possible constraint on the light gravitino mass with
$m_{3/2} < 100$~eV in the light of future precise measurements of
CMB. Although the effects of free-streaming of light gravitino barely
leave an imprint on CMB photons at the time of last scatter, they can
be deflected by the gravitational potential altered after
recombination due to the free-streaming effect, which can be probed
with lensed CMB. Thus we in this paper discussed the effects of the
light gravitino, paying particular attention to the lensing potential,
then investigated the future constraint on its mass. For this purpose,
we adopt the future CMB surveys such as Planck, PolarBeaR and CMBpol
and study the issue by generating posterior distributions with
nested-sampling method.  For a simple (but physically motivated) case,
we obtained the limit on the light gravitino mass, assuming
$m_{3/2}=1$~eV as a fiducial value, as $m_{3/2}\le3.2 \ \mbox{eV} \
\mbox{(95\% C.L.)}$ for the case with Planck+PolarBeaR combined and
$m_{3/2}=1.04^{+0.22}_{-0.26} \ \mbox{eV} \ \mbox{(68\% C.L.)}$ for
CMBpol.  Thus at the time of CMBpol experiment, we can expect that the
(counter-)evidence of the light gravitino can be found in cosmological
observations.

In a simple case, the effective degrees of freedom at the time of the
gravitino decoupling is assumed to have definite value and fixed in
the analysis.  However in some scenarios, this assumption may not
hold, thus we have also made analysis by treating $N_{3/2}$ as a free
parameter.  In addition, in the future precise CMB measurements, the
primordial abundance of $^4$He, which is usually not assumed as a free
parameter but fixed, can also affect the determination of cosmological
parameters, therefore we also performed the analysis by varying $Y_p$
too. When both or one of these parameters are varied, the constraint
on the mass becomes weak.  The results are summarized in Tables
\ref{table:case1}-\ref{table:case4}.
In addition, we also discussed how strong future CMB surveys can 
find an evidence for light gravitino by employing
Bayesian model selection analysis. Even if the mass of gravitino is around 1 eV,
a future CMBpol-like surveys is capable of providing some rather strong evidence
for light gravitino model, which can be even stronger for larger $m_{3/2}$.

In principle the light gravitino mass of $\mathcal O(1)$~eV can be
probed with the LHC experiment with the method proposed in
Ref.~\cite{Hamaguchi:2007ge}.  But this requires some amount of tuning
for the sparticle mass spectrum together with the gravitino mass.
Thus even if the LHC will fail to determine the light gravitino mass,
the future CMB experiments can do this job.

Finally we make some comment on the case where 
massive neutrinos are also included in the analysis.
From atmospheric, solar, reactor 
and accelerator neutrino experiments,  now we know that neutrinos have finite masses.
Furthermore it has been discussed that neutrino masses 
can  be well probed with future CMB survey \cite{Kaplinghat:2003bh,Lesgourgues:2005yv,Perotto:2006rj},  
which motivates us to conduct the analysis 
assuming that neutrinos are massive.
Thus we also investigated the  constraint on 
the light gravitino mass while the mass of neutrino is also varied.
Since the effects of massive neutrino and light gravitino on the lensing potential
are essentially the same, we found that strong degeneracies arise 
in particular, 
between their masses and we could not obtain any meaningful constraints on those.
However, neutrino masses can also be constrained 
in future laboratory experiments 
such as tritium beta-decay
and neutrinoless double beta decay.
Thus in the future, we will also have some inputs from such 
neutrino experiments, which can remove the degeneracy in 
CMB survey. 
In light of these considerations, 
we can expect that cosmology and particle physics 
experiments will push us toward more severe constraint/precise 
determination of the light gravitino mass in the near future.

\bigskip
\bigskip

\noindent 
\section*{Acknowledgment}
The authors would like to thank Kiyotomo Ichiki and Shun Saito for 
useful discussions and providing data to check our numerical calculations.
The authors would also like to thank Oleg Ruchayskiy for useful comments.
This work is supported by Grant-in-Aid for Scientific research from
the Ministry of Education, Science, Sports, and Culture, Japan,
No. 14102004 (M.K.) and No. 19740145 (T.T.), and also by World Premier
International Research Center Initiative, MEXT, Japan.  K.N. and T.S.
would like to thank the Japan Society for the Promotion of Science for
financial support.


\end{document}